\documentclass[showpacs,twocolumn,aps,prb]{revtex4}
\usepackage{graphicx}
\usepackage{amssymb}
\usepackage{amsmath}
\usepackage{bm}
\usepackage{color}

\def\Journal #1,#2,#3,#4#5#6#7{#1 {\bf #2}, #3 (#4#5#6#7)}

\def\GVec#1{\mbox{\boldmath $#1$}}

\begin{document}

\title{Cyclotron resonance of figure-of-eight orbits in a type-II Weyl semimetal \\
}
\author{Mikito Koshino}
\affiliation{Department of Physics, Tohoku University, Sendai 980-8578, Japan}
\date{\today}

\begin{abstract}
We study the cyclotron resonance in the electron-hole joint Fermi surface 
of a type-II Weyl semimetal. 
In magnetic field, the electron and hole pockets touching at the Weyl node 
are hybridized to form quantized Landau levels corresponding to semiclassical 8-shaped orbits.
We calculate the dynamical conductivities for the electric fields oscillating in $x$ and $y$-directions
and find that the resonant frequencies in $x$ and $y$ differ by the factor of two,
reflecting to the figure-of-eight electron motion in the real space.
The peculiar anisotropy in the cyclotron resonance serves as a unique characteristic
of the dumbbell-like Fermi surface.
\end{abstract}



\maketitle

\begin{figure}[h]
\begin{center}
\leavevmode\includegraphics[width=1.\hsize]{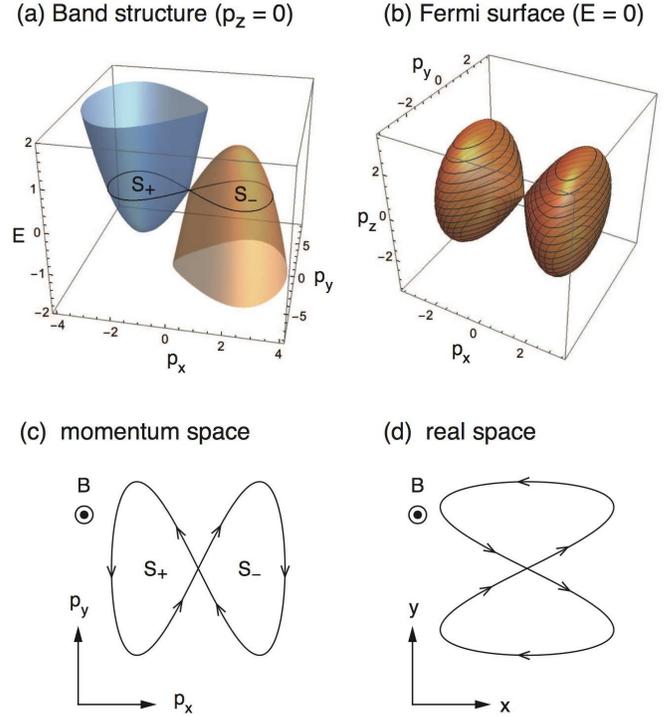}
\end{center}
\caption{Electronic structure of 
the type-II Weyl semimetal described by Eq.\ (\ref{eq_H})
with $w = 2 v$ and $\alpha = 0.1 v/ p_0^2$.
(a) Energy band at $p_z=0$ and 
(b) Fermi surface at $E=0$,
with momentum and energy axes scaled in units of $p_0$ and $v p_0$, respectively.
(c) Semiclassical cyclotron motion at $E=0$ in the momentum space
and (d) in the real space.
}
\label{fig_schem}
\end{figure}

The cyclotron resonance is a fundamental property of metallic systems
and it is also important as a tool to deduce the structure of the Fermi surface.
In the semiclassical picture, an electron under a magnetic field moves along an equi-energy contour
in the momentum space. \cite{ashcroft1976solid}
When the contour is closed, the electron motion becomes periodic
and the dynamical conductivity has a resonant peak at the corresponding frequency. 
Usually the semiclassical orbit is circular and it is classified as either electron-type
 or hole-type depending on the rotating direction.
The recent discovery of so-called type-II Weyl semimetals,
\cite{soluyanov2015type,chang2016prediction,koepernik2016tairte,autes2016robust,xu2016discovery, deng2016experimental,huang2016spectroscopic,liang2016electronic,wang2016spectroscopic,wu2016observation,belopolski2016measuring} on the other hand,
implies a possibility of a unconventional cyclotron orbit as illustrated in Fig.\ \ref{fig_schem},
where the electron pocket and hole pocket are integrated into an ``8-shaped" trajectory.\cite{huhtala2002fermionic,xu2015structured}
The cyclotron resonance in this unusual Fermi surface is the main interest of this study.

A Weyl semimetal is a crystal where the electronic energy bands 
are touching at isolated points (Weyl nodes) in momentum space with a linear dispersion.
\cite{murakami2007phase, burkov2011weyl, burkov2011topological, wan2011topological, yang2011quantum,hosur2013recent,xu2015observation,weng2015weyl,lu2015experimental,xu2015discovery,lv2015experimental,huang2015weyl,xu2015experimental}
The essential feature near a Weyl node is captured by the Hamiltonian 
\begin{align}
H = v \GVec{\sigma}\cdot{\bf p} + ({\bf w}\cdot{\bf p})\sigma_0,
\end{align}
 where $\GVec{\sigma} = (\sigma_x,\sigma_y,\sigma_z)$ are the Pauli matrices
 and $\sigma_0$ is the $2 \times 2$ unit matrix.
The vector ${\bf w}$ describes the tilting of the Weyl cone in  a specific direction.
When $|w/v|<1$, the system is called the type-I Weyl semimetal where the Fermi surface becomes a point
at the band touching energy.
When $|w/v|>1$, it becomes the type-II Weyl semimetal where the electron band and 
the hole band are overlapping in energy, giving an electron-hole joint Fermi surface.
The type-II Weyl semimetal was predicted in various condensed matter systems,\cite{soluyanov2015type,chang2016prediction,koepernik2016tairte,autes2016robust}
and it triggered recent intensive theoretical studies on its electronic properties.
\cite{zyuzin2016anomalous,volovik2016lifshitz,o2016magnetic,
yu2016unusual,udagawa2016field,tchoumakov2016magnetic,khim2016magnetotransport} 
The identification of its peculiar band structure was reported in very recent experiments. \cite{xu2016discovery, deng2016experimental,huang2016spectroscopic,liang2016electronic,wang2016spectroscopic,wu2016observation,belopolski2016measuring}

In this paper, we study the cyclotron resonance of 8-shaped Fermi surface 
of a type-II Weyl semimetal. We consider a system as illustrated in Fig.\ \ref{fig_schem},
and apply a magnetic field in $z$-direction so that the 
both the electron and hole Fermi surfaces are included in the cross-section normal to the field.
We show that, near the Weyl node, the electron and hole pockets are completely hybridized by the magnetic field
to form quantized Landau levels corresponding to semiclassical 8-shaped orbits.
We then calculate the dynamical conductivity for oscillating electric field  in $x$ and $y$-directions
and find that the resonant frequencies in $x$ and $y$ differ by factor of two,
reflecting to the figure-of-eight electron motion in real space.
The peculiar anisotropy in the cyclotron resonance serves as a unique characteristic
of the dumbbell-like Fermi surface.


\begin{figure*}
\begin{center}
\leavevmode\includegraphics[width=0.9\hsize]{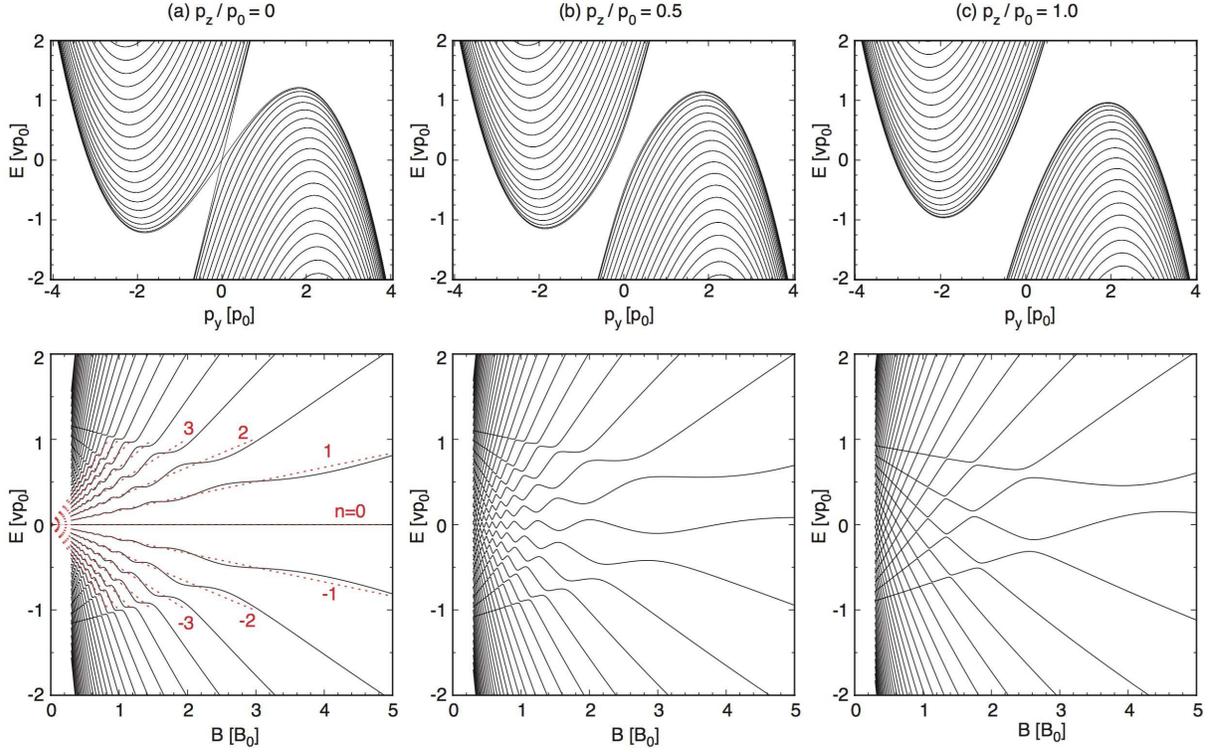}
\end{center}
\caption{(Upper panels)
Band dispersions at the zero magnetic field at different $p_z$'s, in the same model as in Fig.\ \ref{fig_schem}.
(Lower panels) Corresponding Landau level spectra against the field amplitude $B$.
}
\label{fig_spec}
\end{figure*}

We consider the effective Hamiltonian
\begin{equation}
H=v(p_x \sigma_x + p_y \sigma_y + p_z \sigma_z) + (w p_x - \alpha p_x^3) \sigma_0.
\label{eq_H}
\end{equation}
The parameter $w$ describes the tilting of the Weyl cone to $x$-direction,
where the type-II Weyl semimetal is realized in $|w/v|>1$.
The $p_x^3$ term with the parameter $\alpha$ is introduced to close the Fermi surface
within a finite momentum space, as expected in the real materials.
The energy spectrum is given by
\begin{equation}
E = w p_x - \alpha p_x^3 \pm v |\textbf{p}|.
\label{eq_E}
\end{equation}
In the following, we consider a type-II case with $w = 2 v$ and $\alpha = 0.1 v/ p_0^2$, where $p_0$
is the characteristic momentum scale.
Figure \ref{fig_schem}(a) is the actual band structure at $p_z=0$ at this parameter choice,
where the electron band and the hole band overlap in the energy region
and touch at the Weyl node located at $E=0$.
Once $p_z$ is shifted from zero, the band touching is resolved, and 
the 8-shaped Fermi surface is separated into the electron and hole parts.
The whole Fermi surface in three-dimensional space
is plotted in Fig. \ref{fig_schem}(b).

When we apply the magnetic field ${\bf B}=(0,0,B)$ in $z$-direction,
the Hamiltonian is given by Eq.\ (\ref{eq_H}) with 
${\bf p}$ replaced by $\GVec{\pi}=\mathbf{p}+ e\mathbf{A}$,
where $\mathbf{A} $ is vector potential to give $\mathbf{B}=\nabla\times \mathbf{A}$.
The momentum parallel to the field, $p_z$, remains the quantum number.
We define the raising and lowering operators of the Landau levels as
$\pi_x+i\pi_y = \Delta_B a^\dagger$
and $\pi_x-i\pi_y = \Delta_B  a$ with $\Delta_B = \sqrt{2\hbar v^2 e B}$,
which operate on the usual Landau-level wave function
$\phi_n$ in such a way that $a \phi_n = \sqrt{n}\phi_{n-1}$ and
$a^\dagger \phi_n = \sqrt{n+1}\phi_{n+1}$.
The Hamiltonian is then written as
\begin{align}
H &= \Delta_B (a \sigma_+ + a^\dagger \sigma_-)
+ v p_z \sigma_z \nonumber\\
& \quad + w \frac{\Delta_B}{2v} (a+ a^\dagger)\sigma_0
- \alpha  \left(\frac{\Delta_B}{2v}\right)^3 (a+ a^\dagger)^3\sigma_0,
\label{eq_H_B}
\end{align}
where $\sigma_\pm = (\sigma_x \pm i \sigma_y)/2$.

We calculate the Landau levels setting the cut-off in the Landau level index $n_c = 200$.
Figure \ref{fig_spec} presents the band dispersions at the zero magnetic field  at different $p_z$'s (upper panels),
and the corresponding Landau level spectra against the field amplitude $B$ (lower panels),
where the unit of the magnetic field is $B_0=p_0^2/(2e\hbar)$.
 In $p_z/p_0 =1.0$, the electron band and the hole band are well separated in the
 momentum space, and accordingly the spectrum in $B$-field can be viewed as
the independent electron and hole Landau levels crossing each other.
In the large $B$-field, these start to hybridize and exhibit anti-crossing.
In the semiclassical picture, the hybridization effect
is described as a magnetic break down of adjacent orbits.
Since the momentum operators $\pi_x$ and $\pi_y$ yield to the commutation relation
$[\pi_x,\pi_y] = i\hbar eB$, the semiclassical orbit 
has an intrinsic uncertainty in momentum-space position, which is about $\sqrt{\hbar eB}$.
In a rough estimation, the hybridization of adjacent orbits takes place when  the momentum uncertainty exceeds 
the gap in the momentum space, $\delta p$.
In the present unit, the condition is written as $B/B_0 > 2(\delta p/p_0)^2$.

In decreasing $|p_z|$,
the electron and hole orbits get closer
and the level mixing becomes stronger.
At $p_z=0$, the spectrum near zero energy
 is completely reconstructed into a new series of the Landau levels centered at $E=0$.
The reconstructed levels are approximately described by the semiclassical quantization rule
of 8-shaped orbits,
\begin{align}
S_+-S_- = 2\pi \hbar eB n \, \quad (n = 0, \pm1, \pm2,\cdots)
\label{eq_quant}
\end{align}
where $S_\pm=S_\pm(E, p_z)$ is the momentum space area of the electron part and the hole part, respectively,
at the energy $E$ and momentum $p_z$.
The energy levels specified by Eq.\ (\ref{eq_quant}) at $p_z=0$ are indicated by
red dotted lines in Fig.\ \ref{fig_spec}(a), which agree well with the numerical solid curves.
When the electron and hole orbits are far apart, they are separately quantized 
by the usual condition $S_\pm = 2\pi \hbar eB  (n+\gamma)$, where $\gamma$ is 
a constant of the order of unity.

\begin{figure}
\begin{center}
\leavevmode\includegraphics[width=0.8\hsize]{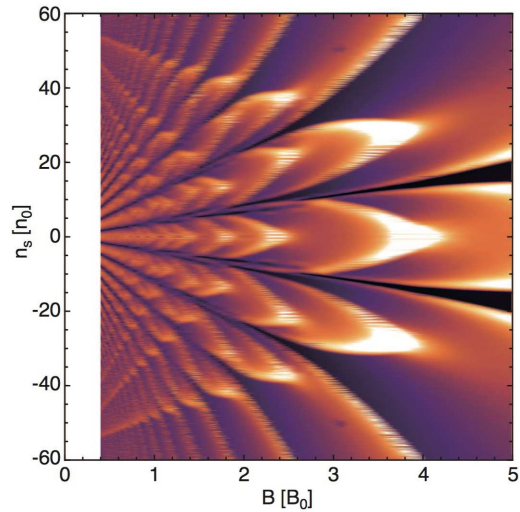}
\end{center}
\caption{Two-dimensional map of the density of states (DOS)
as a function of magnetic field $B$ and the carrier density $n_s$.
Bright color represents high DOS.
}
\label{fig_n-B}
\end{figure}

The reconstruction of the Landau levels directly influences the magnetic oscillation such as de Haas-van Alphen effect.
In Fig.\ \ref{fig_n-B}, we present a two-dimensional map
of the total density of states (DOS) integrated over $p_z$,
in the space of magnetic field $B$ and the carrier density $n_s$.
Here the bright color represents high DOS.
The unit of $n_s$ is taken as $n_0=[p_0/(2\pi\hbar)]^3$ and $n_s=0$ corresponds to the zero Fermi energy.
The change of DOS in moving along a horizontal line (i.e. at fixed $n_s$) simulates the magnetic oscillation
of physical quantities in changing $B$-field.
We clearly recognize a fan structure centered at $n_s=0$, which is
originating from the quantized 8-shaped orbits.
Although the Landau level reconstruction occurs only in the vicinity of $p_z=0$,
it gives a dominant feature in the diagram because the area of the Fermi surface 
takes an extremal value at $p_z=0$. 
The oscillation period of physical quantities against the inverse magnetic field is 
given by $\Delta(1/B) = 2\pi \hbar e/(S_+^0-S_-^0)$ where $S_\pm^0=S_\pm(E, p_z=0)$.
In the diagram, we also see the off-center fans as the secondary feature, 
corresponding to indepedent electron/hole Landau levels.

\begin{figure*}
\begin{center}
\leavevmode\includegraphics[width=0.9\hsize]{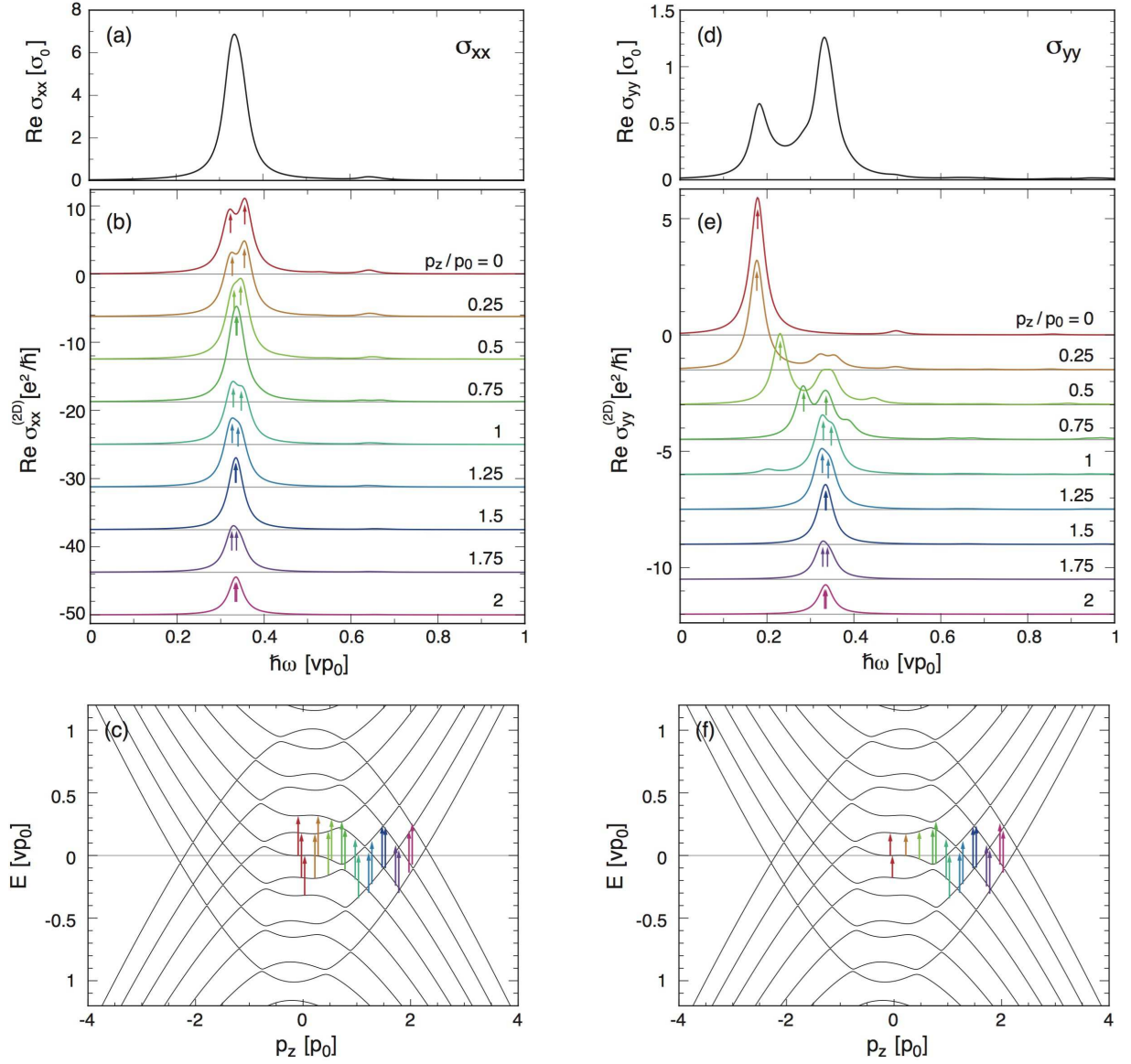}
\end{center}
\caption{Real part of the dynamical conductivity
$\sigma_{xx}$ (left) and $\sigma_{yy}$ (right) at $B=B_0$ and the zero Fermi energy,
In each column, the top figure [(a),(d)] plots the real part of the total conductivity,
the middle [(b),(e)] plots the separate contributions from different $p_z$'s,
and the bottom [(c),(f)] is the  energy levels against $p_z$,
where vertical arrows indicate the excitations corresponding
to the dominant peaks in the middle panel.
}
\label{fig_opt}
\end{figure*}


The Landau level structure can also be probed by the cyclotron resonance.
Here we consider the dynamical conductivity defined by
\begin{align}
\sigma_{\mu\mu}(\omega) = \frac{e^2\hbar}{iV}\sum_{m,n}
\frac{f(\varepsilon_m)-f(\varepsilon_n)}{\varepsilon_m-\varepsilon_n}
\frac{|\langle m | v_\mu |n \rangle|^2}{\varepsilon_m-\varepsilon_n+\hbar\omega + i\delta},
\end{align}
where $\mu = x,y$,  $V$ is the volume of the system,  
$v_\mu = \partial H/\partial p_\mu$ is the velocity operator, 
$\delta$ is the positive infinitesimal, $f(\varepsilon)$
is the Fermi distribution function and $|m \rangle$ and $\varepsilon_m$
are the eigenstate and the eigenenergy, respectively.
For the linearly polarized light with the electric field oscillating in $\mu$-direction, 
the optical absorption is proportional to the real part of $\sigma_{\mu\mu}$.
We numerically calculate of $\sigma_{xx}$ and $\sigma_{yy}$ at $B=B_0$ and the zero Fermi energy,
where the energy broadening effect is included by replacing $\delta$ with the phenomenological constant 
$0.02vp_0$.
The results for $\sigma_{xx}$ and $\sigma_{yy}$ are shown
the left and right columns, respectively, in Fig.\ \ref{fig_opt}.
In each column, the top figure plots the real part of the total conductivity,
and the middle plots the separate contributions from different $p_z$'s.
The bottom presents the energy levels against $p_z$,
where vertical arrows indicate the excitations corresponding to the dominant peaks marked in the middle panel.
The $\sigma_{zz}$ (not shown) is found to be almost negligible compared to $\sigma_{xx}$ and $\sigma_{yy}$,
since the periodic motion of the electrons occurs on $xy$-plane.

The total conductivity is obviously anisotropic,
where $\sigma_{xx}$ has a single peak while $\sigma_{yy}$ exhibits a double-peak structure.
This peculiar feature can be explained by the decomposed spectrum in the middle panel.
In a relatively large $p_z$ (roughly $p_z/p_0 >1$), the Landau levels of electron pocket and the hole pocket
are almost independent and each gives its own cyclotron frequency, 
$\omega_c^\pm = 2\pi e B|\partial S_\pm /\partial E|^{-1}$ as in usual metals.
The energy spacings of the electron and hole Landau levels are given by $\hbar \omega_c^\pm$, respectively.
In the present model, $\omega_c^+$ and $\omega_c^-$ nearly coincide, so that we have
a single peak at the same position in $\sigma_{xx}$ and  $\sigma_{yy}$.

At $p_z =0$, on the other hand, the electron and hole pockets are united into an 8-shaped orbit.
The time period to complete the figure-of-eight motion is given by $2\pi/\omega_c^{\infty}$ where
\begin{align}
\omega_c^{\infty} = 2\pi e B\left|\frac{\partial}{\partial E} (S_+ - S_-) \right|^{-1},
\end{align}
and the spacing of the reconstructed Landau level [Eq.\ (\ref{eq_quant})] becomes $\hbar \omega_c^{\infty}$.
The important point is that the optical resonance does not always occur at $\omega=\omega_c^{\infty}$
but depends on the electric-field direction: $\sigma_{xx}$ peaks at $2\omega_c^{\infty}$
while $\sigma_{yy}$ at $\omega_c^{\infty}$, as seen in Fig.\ \ref{fig_opt}.
The reason for this can be traced back to the figure-of-eight electron motion
shown in Fig.\ \ref{fig_schem}(c).
The real-space trajectory is approximately described by a Lissajous curve
\begin{align}
x(t) = -\sin 2 \omega_c^{\infty} t, \quad y(t) = \cos \omega_c^{\infty} t,
\end{align}
where the electron oscillates in $x$-direction twice as frequently as in $y$,
leading to the factor-of-two difference in the resonance frequencies in $\sigma_{xx}$ and $\sigma_{yy}$.
The optical resonance in $\sigma_{xx}$ 
corresponds to the excitation from $n$ to $n+2$ in the Landau levels [Eq.\ (\ref{eq_quant})],
and that in $\sigma_{yy}$ to the excitation  from $n$ to $n+1$, as shown in the bottom panels of Fig.\ \ref{fig_opt}. 
The slightly-shifted double-peak structure observed in $\sigma_{xx}(p_z =0)$
is due to the deviation from the equal spacing in the reconstructed Landau levels,
which gives a small difference between the excitation energy from $n=-1$ to $1$
and that from $n=-2$ to $0$ (and 0 to 2).

Here $\omega_c^{\infty}$ is nearly half of $\omega_c^+ \sim \omega_c^-$,
because it takes twice as long time to complete 8-shaped motion than to round the individual electron or hole part.
When integrating the dynamical conductivity over $p_z$, therefore,
the total $\sigma_{xx}$ end up with a single peak near $2\omega_c^{\infty}\sim \omega_c^\pm$,
while the $\sigma_{yy}$ exhibits separated two peaks at $\omega_c^\pm$ and at $\omega_c^{\infty}$.



To summarize, we calculated the Landau level spectrum and the optical absorption
in an effective continuum model for the type-II Weyl semimetal.
The characteristic figure-of-eight cyclotron motion leads to 
the Landau level rearrangement, and a strong anisotropy in the cyclotron resonance,
where the resonant frequency differs by factor two depending on the light polarizing direction.

The author would like to thank Tetsuro Habe and Riichiro Saito for useful discussions. 
This work was supported by JSPS Grants-in-Aid for Scientific research (Grants No. 25107005).

{\it Note added:}
During the completion of this manuscript, we became aware of a recent theoretical work,
which also reports the magnetic breakdown and Landau level formation
in the type-II Weyl semimetal.  \cite{o2016magnetic}.

\bibliography{weyl_typeII}

\end{document}